\documentclass[aps,prl,superscriptaddress,reprint]{revtex4-1}

\usepackage{graphicx}
\usepackage{amsmath}

\begin{document}

\title{Rapid production of uniformly-filled arrays of neutral atoms}

\author{Brian J. Lester}
\email[E-mail: ]{blester@jila.colorado.edu}
\affiliation{JILA, National Institute of Standards and Technology and University of Colorado, and
Department of Physics, University of Colorado, Boulder, Colorado 80309, USA}
\author{Niclas Luick}
\affiliation{JILA, National Institute of Standards and Technology and University of Colorado, and
Department of Physics, University of Colorado, Boulder, Colorado 80309, USA}
\affiliation{Institut f\"{u}r Laserphysik, Universit\"{a}t Hamburg, Luruper Chaussee 149, 22761 Hamburg, Germany}
\author{Adam M. Kaufman}
\affiliation{JILA, National Institute of Standards and Technology and University of Colorado, and
Department of Physics, University of Colorado, Boulder, Colorado 80309, USA}
\author{Collin M. Reynolds}
\affiliation{JILA, National Institute of Standards and Technology and University of Colorado, and
Department of Physics, University of Colorado, Boulder, Colorado 80309, USA}
\author{Cindy A. Regal}
\email[E-mail: ]{regal@colorado.edu}
\affiliation{JILA, National Institute of Standards and Technology and University of Colorado, and
Department of Physics, University of Colorado, Boulder, Colorado 80309, USA}

\begin{abstract} 
We demonstrate rapid loading of a small array of optical tweezers with a single $^{87}$Rb atom per site. We find that loading efficiencies of up to 90\% per tweezer are achievable in less than 170~ms for traps separated by more than $1.7 \,\mu$m. Interestingly, we find the load efficiency is affected by nearby traps and present the efficiency as a function of the spacing between two optical tweezers. This enhanced loading, combined with subsequent rearranging of filled sites, will enable the study of quantum many-body systems via quantum gas assembly.
\end{abstract}

\date{June 12, 2015}

\maketitle

A frontier in atomic physics is the study of quantum many-body physics on a microscopic scale. Recent experiments have shown the power of microscopy of degenerate quantum gases in optical lattices~\cite{Bakr2009,Sherson2010}. An exciting prospect is not only imaging quantum gases, but assembling them into a well-known initial configuration from single-atom building blocks and then observing the resulting dynamics with single-atom resolution.  Wavelength-scale optical dipole traps, or optical tweezers, are an attractive platform for control of neutral atoms because they allow repositioning of the atoms after state preparation and site-resolved imaging.  Using optical tweezers, long-range interactions between neutral atoms have been harnessed via Rydberg blockade \cite{Wilk2010,Isenhower2010,Barredo2015}, and it is now possible to observe controlled interactions and interference between bosonic and fermionic atoms placed individually in their motional ground state~\cite{Kaufman2012,Kaufman2014,Murmann2015}.  While optical tweezer traps can be scaled to arrays~\cite{Bergamini2004,Zimmermann2011,Piotrowicz2013,Nogrette2014}, realizing an ordered array with a single atom per trap is difficult and is a problem of long-standing interest~\cite{Weiss2004,Miroshnychenko2006,Beugnon2007,Grunzweig2010}.

Early experiments with optical tweezers demonstrated sub-Poissonian atom-number statistics using light-assisted collisions that rapidly expel pairs of atoms, a process known as collisional blockade~\cite{Schlosser2001,Schlosser2002,Fuhrmanek2012,Sompet2013}. This has become a reliable method to isolate single atoms, as well as the basis for parity imaging in quantum-gas microscopes~\cite{Schlosser2001,Bakr2009,Sherson2010}. However, the collisional blockade also limits loading efficiencies to approximately 50\%, making the probability to uniformly-fill large arrays prohibitively small~\cite{Schlosser2002,Fuhrmanek2012, Sompet2013}.

\begin{figure}[tb!]
	\begin{center}
		\includegraphics[width=\columnwidth]{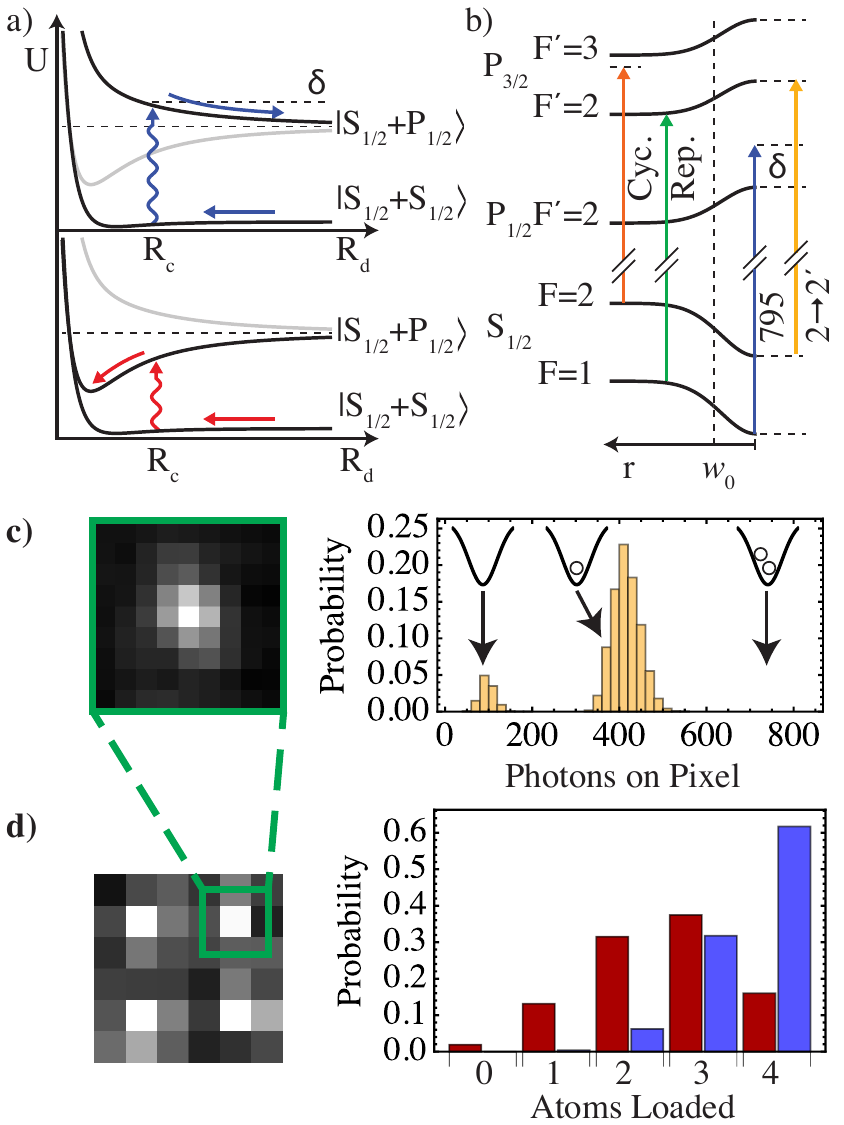}
		\caption{(color online).~\emph{Enhanced loading and beam schematic.} 
			(a)~Light-assisted collisions for blue-detuned light (top) and red-detuned light (bottom).  As the atoms approach R$_C$ in the ground-state $|S_{1/2}+S_{1/2}\rangle$ potential, the light becomes resonant with excitation to the repulsive (attractive) $|S_{1/2}+P_{1/2}\rangle$ molecular potential.
			(b)~Level diagram showing the relevant levels of $^{87}$Rb and the free-space (left) or trap-shifted (right) frequencies required for enhanced loading (level spacings not to scale). 
			(c)~An average of many images of a single tweezer at full resolution.  Histogram showing the number of photons detected on the highlighted binned-pixel after the 155~ms enhanced loading procedure.  
			(d)~A single image of four atoms in an array of four optical tweezers separated by $4.18\,\mu$m along each dimension. Comparison of loading probability in an array of four tweezers for loading using PGC (red bars) and the enhanced loading procedure (blue bars). 
		} 
		\label{loadingAndSchematic}
	\end{center}
\end{figure}

Careful studies of light-assisted collisions in optical dipole traps hold promise for realizing deterministic loading of arrays of atoms~\cite{Grunzweig2010,Carpentier2013}. Light-assisted collisions are successfully described by transitions between molecular potentials that become resonant with the light at specific interatomic separations, R$_{C}$ [Fig.~\ref{loadingAndSchematic}(a)]~\cite{Weiner1999}. In the case of light that is red-detuned from the bare atomic transition, the atoms associate to an attractive potential and can gain a large kinetic energy, leading to loss of both atoms from the trap. Conversely, when the light is blue-detuned, the atoms associate to a repulsive potential where the maximum kinetic energy gained is set by the detuning~\cite{Hoffmann1996}. This control has been used to preferentially expel single $^{85}$Rb atoms from a trap, enabling the isolation of single atoms with high probability~\cite{Grunzweig2010,Carpentier2013}. However, open questions have been whether this technique can be extended to efficiently load arrays of traps or be adapted to atomic species with a different hyperfine structure.

Here, we report near-deterministic loading of single $^{87}$Rb atoms that is more rapid than previous realizations and show that this loading can be used to prepare a uniformly-filled array of traps.  This technique realizes a $2\times2$ array of atoms in optical tweezers in more than 60\% of experimental runs [Fig.~\ref{loadingAndSchematic}(d)].  Further, we study how close two optical tweezers can be loaded without deleterious effects.  To enable the creation of closely-spaced atom arrays, we use tight optical tweezers that have a volume more than an order of magnitude smaller than was used in previous work. This causes the collisional dynamics to occur on a much faster time scale~\cite{Schlosser2002}, which prevents initially trapping many atoms in a single tweezer.  Hence, in our method a set of collisional beams are applied in conjunction with loading from a magneto-optical trap (MOT).  Despite this difference, we find the optical detunings and powers required are similar to previous work with $^{85}$Rb~\cite{Grunzweig2010,Carpentier2013}.

The experimental apparatus generates optical tweezers by focusing $\lambda=852$ nm light to a $1/e^2$ radius of $w_0=0.71\, \mu$m~\cite{Shih2013,LeKien2013,Kaufman2014}. Arrays of optical tweezers are created by generating multiple deflections in each of two acousto-optic modulators (AOMs) oriented to generate perpendicular deflections. In our MOT, we combine a magnetic field gradient with three beams that are each retro-reflected in the  $\sigma^+ - \sigma^-$ polarization configuration and contain ``cycling'' light, which is 25~MHz red-detuned of the free-space D2~$F=2 \rightarrow F^{\prime}=3$ transition [Fig.~\ref{loadingAndSchematic}(b)]; two of the beams also contain ``repump'' light, which is resonant with the free-space D2~$F=1 \rightarrow F^{\prime}=2$ transition. 

In our enhanced loading procedure, our optical tweezers are overlapped with the MOT along with two ``collisional beams'' [Fig.~\ref{loadingAndSchematic}(b)]: The ``795 beam'' is blue-detuned from the trap-shifted D1~$F=1 \rightarrow F^{\prime}=2$ transition and drives the blue-detuned light-assisted collisions. The ``$2-2^\prime$'' beam is near resonant with the trap-shifted D2~$F=2 \rightarrow F^{\prime}=2$ transition and quickly pumps trapped atoms to the $F=1$ manifold.  After loading the MOT for 110-135~ms with the collisional beams on, we turn off the magnetic-field gradient and zero the magnetic field, while keeping only the collisional beams on for an additional 35~ms, which ensures there are never two atoms in the trap after the loading procedure. To give the colliding atoms just enough kinetic energy for one atom to escape from the trap, the 795 beam should be detuned by approximately the trap depth (in units of $h$). We find it is important to have a large trap depth ($h\times 73$~MHz for these data) to overcome single-particle loss due to higher scattering rates at smaller 795 beam detunings.

Atoms are detected via the fluorescence collected during 25~ms of polarization-gradient cooling (PGC) in the trap~\cite{Lester2014}. The histogram in Fig.~\ref{loadingAndSchematic}(c) depicts the number of photons detected on the highlighted pixel in Fig.~\ref{loadingAndSchematic}(d) over 2000 runs of the experiment. The two peaks in the histogram correspond to runs with zero (100 photons) and one (430 photons) atom in the trap, indicating one atom in (88.7$\pm$0.4)\% of runs. There are never two atoms present in the trap, as indicated by the lack of images with more than 550 photons detected. The collisional beam parameters used to achieve this result are given in Table~\ref{beamParameters}. For comparison, single atoms can be loaded into our $h\times 23$~MHz depth trap via PGC in 63\% of attempts, as shown by the red bars in Fig.~\ref{loadingAndSchematic}(d)~\cite{Kaufman2012,Sompet2013,Lester2014}. 

\begin{table}[tb!]
	\begin{center}
	\begin{tabular}{l || c| c}
		Parameter              & 795 Beam & 2-2$^{\prime}$ Beam \\
		\hline
		Detuning 	& \quad$\delta=+50.2$~MHz \quad 	&\quad +8.6 MHz \quad \\
		Intensity	& 40 $\frac{\text{mW}}{\text{cm}^2}$	& 14  $\frac{\text{mW}}{\text{cm}^2}$\\
		Polarization \quad	& linear ($\pi$) &	\quad lin$\perp$lin (unpolarized)
	\end{tabular}
	\caption{
	Collisional beam parameters used for loading the $2\times2$ array of optical tweezers shown in Fig.~\ref{loadingAndSchematic}(d). Representative saturation intensities for these transitions are between 3 and 5 $\frac{\text{mW}}{\text{cm}^2}$ \cite{Steck2010}. 
	}
	\label{beamParameters}
	\end{center}
\end{table}

\begin{figure}[tb!]
	\begin{center}
		\includegraphics[width=\columnwidth]{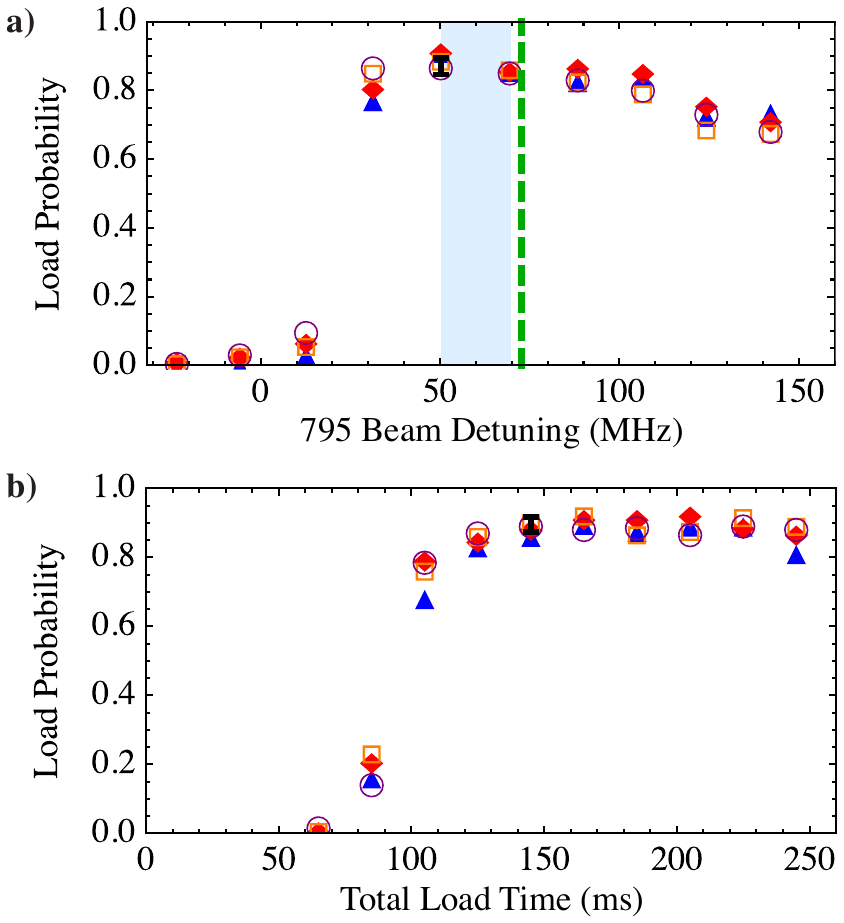}
		\caption{(color online).~\emph{Loading an array of optical tweezers.} 
		(a)~Loading probability as a function of the 795 beam detuning for each of the four wells in the 2$\times$2 array shown in Fig.~\ref{loadingAndSchematic}(d); each shape corresponds to data from one of the four wells, shown individually to demonstrate the consistency. The green dashed line is the trap depth (in units of $h$).  The light-blue band represents the range of detunings used for enhanced loading. 
		(b)~Loading probability as a function of the total length of the enhanced loading procedure. Notice the rapid increase in the load probability as the MOT density increases, saturating around 90\% in under 170~ms. All of the data points correspond to the loading probability from 200 repetitions of the loading procedure; we show a single representative error bar (standard error in the measurement), which is indicated in black, on the fifth data point in each plot. 
		}
		\label{parameterScans}
	\end{center}
\end{figure} 

The dependence on the 795 beam detuning is shown in Fig.~\ref{parameterScans}(a).  The detuning is calibrated relative to the trap-shifted transition, which is approximately 115~MHz blue of the free-space resonance. Near-resonance, the large scattering rate causes rapid single-atom loss and the load probability drops. The light-blue band depicts the range of detunings used for enhanced loading, while the green dashed line represents the trap depth (in units of $h$).  The use of detunings smaller than the trap depth is expected. It is unlikely for both atoms to be at the bottom of the trap prior to colliding and thus less energy is required to remove one atom.  The loading efficiency decreases at larger detunings because the energy gained becomes large enough for both atoms to be kicked out in a single collision. The $2-2^\prime$ beam should be resonant with the trap-shifted transition, but the loading efficiency is not particularly sensitive to the exact detuning except for a rapid drop-off (due to single-particle heating) as the beam is tuned red of the transition. The $+8.6$~MHz detuning (Table~\ref{beamParameters}) serves as a buffer against slow drifts in the light-shifted resonance without degradation in the peak loading efficiency.  However, we find that the beam intensities required for both of the collisional beams are significantly higher than one might expect, i.e., well above the saturation intensity for $^{87}$Rb. For example, we observe that reducing the intensity of either beam by a factor of 4 reduces the load probability to 70\% or less.

We found that the polarization of the collisional beams did not have a  significant effect on the loading efficiency, but we indicate those used for our data in Table~\ref{beamParameters}~\cite{Carpentier2013}. Because there is no quantization axis during the loading, the polarization is not well-defined with respect to the atoms. Hence, we define $\pi$-polarization to be linearly polarized along the optical tweezer axis (the axis of propagation of the trapping light), and this is the polarization used for the 795 beam. The $2-2^\prime$ beam is retro-reflected in a lin$\perp$lin configuration, with the beams crossing the traps at roughly 45$^\circ$ to the tweezer axis, giving projections onto all polarizations. We observe that the loading efficiency is independent of whether the $2-2^\prime$ light is retro-reflected or in a single-pass configuration. This indicates that the $2-2^\prime$ light is not cooling the atoms via the lin$\perp$lin polarization configuration.

Finally, we show the loading efficiency as a function of the total length of the loading procedure, which starts when the light is first turned on and ends before imaging, in Fig.~\ref{parameterScans}(b). The load time here is likely mainly limited by the time required for the MOT to achieve a sufficient density, as evidenced by our observation of similar required time scales for typical loading~\cite{Kaufman2012}. And hence, it is likely that improving the MOT loading rate (e.g., by using a different beam geometry) would shorten the total loading time.

\begin{figure}[tb!]
	\begin{center}
		\includegraphics[width=\columnwidth]{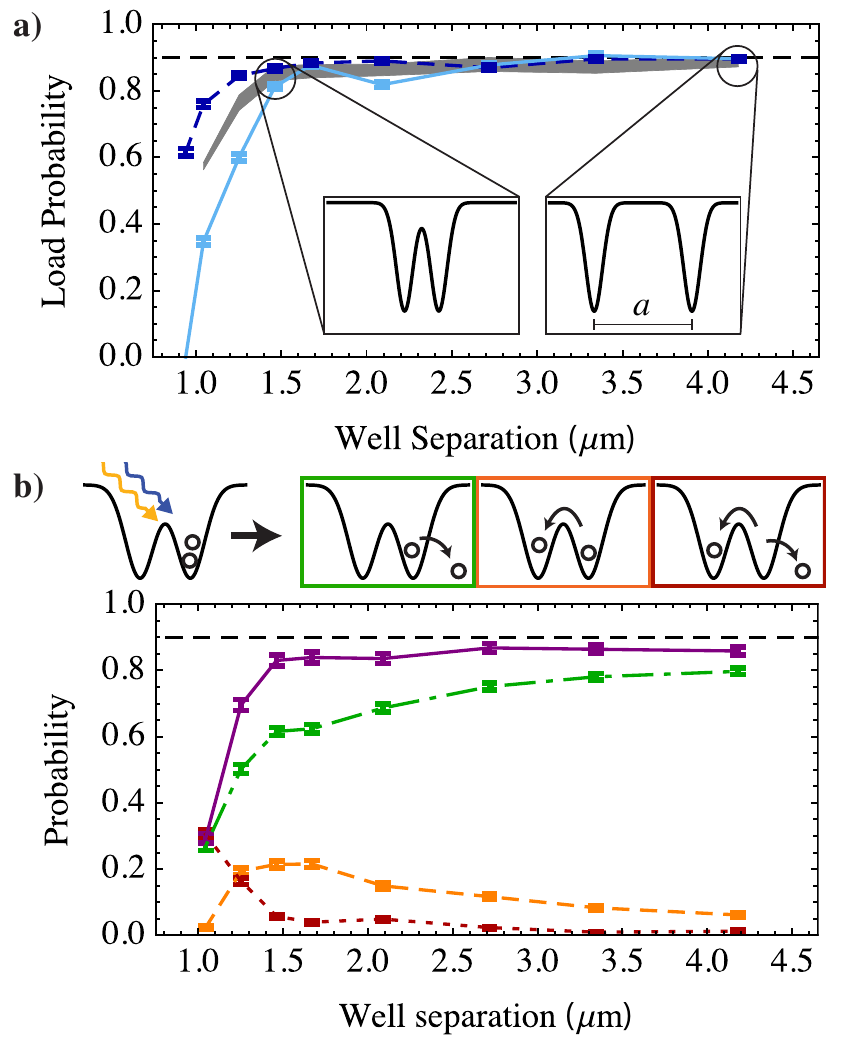}
		\caption{(color online).~\emph{Effect of tweezer spacing on loading.} 
		(a)~Maximum loading probability achieved (per tweezer) after optimizing the loading parameters at each well spacing (dark-blue dashed line)  and the loading probability when using the isolated-tweezer beam parameters at each spacing (light-blue solid line) versus the spacing between the center of two tweezers $a$. The loading probability calculated from a Monte-Carlo simulation of the loading process (gray band) using the measured probabilities of the four possible outcomes (three of which are shown in part b) during each collision event and neglecting any single-particle effects.
		(b)~Controlled experiment studying the effect of collisional light on two trapped atoms in proximity to a second trap. Measured probabilities of a single atom remaining in the right well (dash-dotted green line), one atom remaining in each well (dashed orange line), a single atom remaining in the left well (dotted red line), and the total probability for an atom to remain in the right well (solid purple line) as a function of the spacing between the optical tweezers. 
		}
		\label{spacingData}
	\end{center}
\end{figure}

When applying the enhanced loading protocol to a tweezer array, we can observe a reduction in the load efficiency of each tweezer. To study this effect, we measure the loading efficiency as a function of the spacing between two optical tweezers $a$ [Fig.~\ref{spacingData}(a)]. The points connected by the solid light-blue line depict the measured load efficiency using collisional beam parameters optimized for tweezers at a separation of $a=4.18 \, \mu$m; these isolated-tweezer beam parameters are the same as those given in Table~\ref{beamParameters}, except with a larger 795 beam detuning of $\delta=+ 69.4$~MHz. The points connected by the dashed dark-blue line indicate the highest loading efficiencies achieved by optimizing the collisional beam parameters at each tweezer separation. As a guide to the eye, the dashed black line indicates a loading probability of 90\%. The insets depict the total potential of two gaussian tweezers with $w_0=0.71 \,\mu$m and $a=1.46\, \mu$m and  $4.18\,\mu$m.  Notice that the reduced load efficiency occurs for values of $a$ where the barrier between the two wells is lowered.

To gain further insight into the reduced loading probability, we perform a separate experiment where we initialize two atoms in the right tweezer (by combining two traps with a single atom each, as determined by post selection). We apply the collisional beams to the pair of atoms for 35~ms, which should be sufficient time to ensure that the atoms do not remain in the same trap, but also means that multiple light-assisted collisions could occur. This experiment allows us to observe the resulting trap occupancies after collisions between two atoms in the presence of a nearby tweezer. The cartoons above the plot in Fig.~\ref{spacingData}(b) depict measured outcomes: A single atom remaining in the right well, one atom in each well, and a single atom remaining in the left well. In addition, two-atom loss can occur. The solid purple line is the sum of the first two, which both result in a single atom in the right well; this is the closest analog to load efficiency.

Using the outcome probabilities from Fig.~\ref{spacingData}(b), and assuming the effects of collisions that do not remove an atom are negligible, we perform a simple Monte-Carlo simulation to estimate the expected loading efficiency. The results of this simulation [gray band in Fig.~\ref{spacingData}(a)], exhibit the same character as the measured loading data. The importance of this simulation is that it demonstrates that atoms moving between the wells can affect the final loading probabilities. We also see that the numbers it gives are reasonable: The simulated load efficiency is between the loading measured with the same collisional beam parameters (solid light-blue line) and the highest achievable loading rate (dashed dark-blue line) and the addition of the magnetic field gradient and MOT light during load could alter the outcome probabilities. While we can not exclude the possibility that the reduced loading efficiency is due to proximity between the atoms themselves (e.g., due to expelled atoms colliding with atoms in nearby traps), our data are consistent with this reduction being solely due to the deformation of the combined potential leading to a higher probability of atoms moving between wells. Thus, it is possible that atoms could be efficiently loaded into potentials separated by less than $1.7\,\mu$m, provided the potential barrier between neighboring traps is large enough that they are effectively independent traps.

The loading procedure presented already makes larger-scale experiments in optical tweezers feasible, especially when combined with trap rearrangement based on occupancy~\cite{Weiss2004,Miroshnychenko2006}.  But it is still relevant to ask what would improve this loading efficiency. Using a larger trap depth will help mitigate two important sources of loss: Single-atom heating from scattering of the 795 beam and pair loss due to red-detuned collisions from the MOT light. However even with larger trap depths, more consistent initial conditions are necessary to achieve the highest possible load efficiencies~\cite{Carpentier2013}. It may be possible to use a form of blue-detuned Sisyphus cooling as a mechanism to cool single trapped atoms that does not interfere with the primary light-assisted collisions in the presence of a second atom~\cite{Wineland1992}. 

In conclusion, we have demonstrated a loading procedure that allows for the rapid preparation of uniformly-filled arrays of single neutral atoms and only requires the addition of two collisional beams. We observed a reduction in the loading efficiency for optical tweezers in close-proximity, but found that, for isolated potential wells, we achieve up to 90\% loading efficiency per well. This procedure will not only enable the creation of uniform atom arrays via optical tweezers, but could also be applicable to the loading of optical lattices or even nanophotonic structures~\cite{Hung2013,Tiecke2014}.
\\

\begin{acknowledgments}
We thank Mikkel Andersen and his colleagues for a helpful discussion, as well as Mike Foss-Feig, Yiheng Lin, and Randall Ball for useful comments.  This work was supported by the David and Lucile Packard Foundation and the National Science Foundation under grant number 1125844.  CAR acknowledges support from the Clare Boothe Luce Foundation. NL acknowledges support from the Studienstiftung des deutschen Volkes.
\end{acknowledgments}


%

\end{document}